\documentstyle[aps,prb,psfig,twocolumn]{revtex}
\begin{document}
\draft
\wideabs{
\title{Anomalous Larmour Frequency Dependence of Proton Spin-Lattice Relaxation Time (T$_1$) in the Ferroelectric Glycine Phosphite}
\author{ R. Kannan {\footnote{rkannan@physics.iisc.ernet.in}}, K. P. Ramesh{\footnote {kpramesh@physics.iisc.ernet.in}}and J. Ramakrishna {\footnote{corresponding authors Tel:+090-+080-3092722,Fax:+090-+080-3461602, email:jr@physics.iisc.ernet.in or kpramesh@physics.iisc.ernet.in}}}
\address{Department of Physics, Indian Institute of Science, Bangalore-560012,India}
\date{8 July 2000}
\maketitle
\begin{abstract}
 We report here the results of $^1$H NMR spin-lattice relaxation time (T$_1$) studies in Glycine phosphite which is a ferroelectric below 224 K. The experiments have been carried out in the temperature range from 200 K to 419 K and at two Larmour frequencies of 11.40MHz and 23.56 MHz.  We have  noticed a Larmour frequency dependence  on the high temperature side of  the T$_1$ minimum . A model is proposed based on the BPP theory to explain the observation.
\end{abstract}
}
 \section{Introduction}
Glycine phosphite (abbreviated as GPI) is one of  the many inorganic addition compounds of amino acid group. Most of these addition  compounds show interesting physical properties. Some well known ferroelectric addition compounds formed with the amino acid Glycine ($^+$NH$_3$CH$_2$COO$^-$) are tri glycine sulphate (TGS), selenate (TGSe), tetrafluoroberrylate (TGFBe) \sloppy and diglycine nitrate (DGN) [1-4]. Another interesting group of amino acid addition compounds, which is formed by the $\alpha$ amino acid betaine ((CH$_3$)$_3$N$^{+}$CH$_{2}$COO$^{-}$) includes among others,  betaine phosphate (BP) which is an antiferroelectric, betaine phosphite (BPI) and betaine arsenate (BA) which are ferroelectrics and betaine calciumchloride dihydrate (BCCD) which exhibits a number of phases including incommensurate ones [5-8]. 

GPI (also the other inorganic addition compounds mentioned above) belongs to the family of hydrogen bonded crystals. At room temperature it  is monoclinic with space group P2$_1$/a ~\cite{Av93}.  The structure consists of  a  layered arrangement of planes containing infinite [H(HPO$_3$)]$_n^{n-}$ chains, alternating with planes built by the organic cations, both layers developing parallel to the {\bf bc} plane. The chains of  HPO$_3$ tetrahedra which are connected by strong hydrogen bonds are directed along the {\bf c}-axis. 

GPI undergoes a ferroelectric phase transition at about 224 K ~\cite{Da96}. The spontaneous polarisation appears along the {\bf b} direction which  is perpendicular to the hydrogen-bonded chains of the phosphite anions. This is not so in the case of BPI or caesium dihydrogen phosphate where the hydrogen-bonded chains are parallel to the ferroelectric axis ~\cite{Ab88,Ue76}. However, in ferroelectric potassium dihydrogen phosphate (KDP),  the  hydrogen bonds are perpendicular to the ferroelectric axis. 

Dielectric, infrared, Raman, EPR and DSC measurements have indicated that the ferroelectric phase transition in GPI is second order and of order-disorder type ~\cite{Ba96,Mo98,Da96}.  Tritt {\it et.al.}~\cite{Tr98} have performed proton NMR spin-lattice relaxation time measurements at 90 MHz to understand  the role played by glycinium cations  in the ferroelectric transition in GPI . The main feature  in the variation of spin-lattice relaxation time (T$_1$) with temperature is the  T$_1$ minimum observed at 294 K.

We have taken up the $^1$ H  NMR spin-lattice relaxation time measurements in GPI as a part of our investigation to study proton dynamics in the ``Orientational glass'' state exhibited by the mixed systems of glycine phosphite and betaine phosphate (abbreviated as BP). Surprisingly, the spin-lattice relaxation time versus 1000/T ( K$^{-1}$) plot in GPI has exhibited a  frequency dispersion in the high temperature side of the T$_1$ minimum. We present here our experimental results and a  model  to explain the observed T$_1$ dispersion.

\section{Experimental Details}
Polycrystalline samples of  GPI were grown at 308 K from aqueous solution containing glycine and phosphorous acid in the ratio 1:1.by slow evaporation. The samples were purified by recrystallisation. The d-spacings obtained using the  powder X-ray diffraction technique agree well with the published  crystal structure data ~\cite{Av93}. The crystal structure as mentioned earlier, is monoclinic with space group P2$_1$/a and the unit cell dimensions are a = 0.9792nm, b = 0.84787nm, c = 0.7411nm, $\beta$ = 100.43  \AA \sloppy  and  Z = 4.

 For our experiments, the polycrystalline sample  was  powdered and sealed   in a glass tube after evacuation. $^1$ H NMR spin- lattice relaxation time were measured at two Larmor \sloppy frequencies of 11.40MHz and 23.56MHz using a home made pulsed NMR spectrometer working in the range 3-30MHz.  A $\pi-\tau-\frac{\pi}{2}$  pulse sequence was employed for spin-lattice relaxation time measurements. The recovery of magnetization followed a single exponential in the temperature range studied . Above 419K the sample melts. The melting temperature observed matched with that obtained from DSC experiments. The temperature of the sample was controlled using a nitrogen gas flow  arrangement and measured using a Pt-100 sensor. 

\section{Results and Discussion} 
Fig:1 shows a  plot of spin-lattice relaxation time (T$_1$)  with 1000/T at the two Larmour frequencies of  11.40 MHz and 23.56 MHz. T$_1$ passes through a minimum as the temperature is decreased. The minimum in T$_1$ is at T = 262K  (1000/T = 3.82 K$^{-1}$) and has a value of  10.8 mS at 11.40 MHz.  At 23.56 MHz the T$_1$ minimum occurs at 276 K (1000/T = 3.62 K$^{-1}$) and has a value of 21.7 mS. An interesting feature seen  in  Fig:1 is the Larmour frequency dependence of  T$_1$  at temperatures above the T$_1$ minimum till the highest temperature ( 419 K). The  T$_1$ increases with decreasing temperature below the T$_1$ minimum till the lowest temperature studied.  The T$_1$ dispersion on the low temperature side of the T$_1$  minimum is as expected  from the BPP theory ~\cite{Bl48}. 

In the earlier $^1$H NMR spin-lattice relaxation work in the same system by Tritt {\it et.al.}~\cite{Tr98}\linebreak at a single Larmour frequency of 90 MHz, the spin-lattice relaxation behaviour has been explained by assuming that the reorientation of the  NH$_3^{+}$  relaxes all the protons via spin-diffusion in the entire temperature range from 195K to 420 K. However, the model used by Tritt {\it et.al.} could not explain  our T$_1$  results, especially the T$_1$  dispersion observed on the high temperature side of the T$_1$ minimum.  The expression for the proton spin-lattice relaxation rate used by Tritt {\it et.al.} ~\cite{Tr98} is given as,
\begin{equation}
\frac{1}{T_{1NH_3}} = \frac{3}{8}D \left ( \frac{\tau_c}{1+\omega_0^{2}\tau_c^{2}} +\frac{4\tau_c}{1+4\omega_0^{2}\tau_c^{2}}\right ) ,
\label{T1}
\end{equation} 
where  $\tau_c  =  \tau_{c0}  exp\left (\frac{E_c}{kT}\right )$ and $D  =  \frac{9}{20}\left ( \frac{\mu_0}{4\pi}\right )^2 \gamma^4 \hbar^2 r^{-6}$. The factor $\frac{3}{8}$ in the Eqn:\ref{T1} arises  from  spin -diffusion and in this particular case there are three relaxing  protons belonging to the NH$_3^{+}$  group, relaxing the total number of eight protons in the molecule. $\tau_c$ is the correlation time of  the NH$_3^{+}$  group reorientation which is assumed to follow the Arrhenius behaviour, E$_c$ is the activation energy and D is the relaxation constant for the  NH$_3^{+}$  group reorientation. The Tritt model predicts a Larmour frequency independent behaviour for  T$_1$ on the high temperature side of  T$_1$ minimum . The T$_1$ minimum due to reorientation of the  NH$_3^{+}$  group alone without spin-diffusion will be  5.5 mS and 11.4 mS  at the two frequencies of our study and these are half of the values that have been observed. This can occur if there are  inequivalent reorienting groups in the system contributing to relaxation ~\cite{Ve93,Ni81}. It is known that if there are N$_a$ number of ``a'' type groups and N$_b$ number of ``b'' type groups, then the relaxation rate due to these inequivalent groups is given as,
\begin{equation}
\frac{1}{T_1} = \left ( N_a \frac{1}{T_{1a}}+ N_b \frac{1}{T_{1b}}\right ) \frac{1}{N_a+N_b}
\label{inequivalent}
\end{equation}
 We have fitted our data at 11.40 MHz and 23.56 MHz to Eqn:\ref{inequivalent}  and the best fit  T$_1$  values do not agree with the observed values for both the cases of  N$_a$ = N$_b$ and N$_a$$\not =$N$_b$. We have also considered the possibility of inequivalence with spin-diffusion but in vein. These failed attempts to fit the data to inequivalent glycinium cations agree with the observations from the vibrational studies in the paraelectric state of GPI ~\cite{Ba96}. 

\begin{figure}
\psfig{file=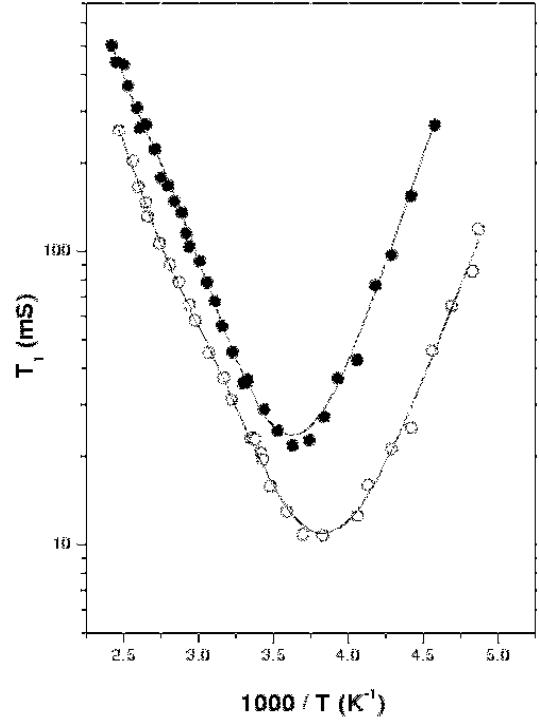,width=8cm}
\caption{ $^1$H NMR spin-lattice relaxation time in GPI as a function of temperature. Open circles correspond to 11.40 MHz and solid circles correspond to 23.56 MHz. Lines denote the theoretical fit to the corresponding data.}
\end{figure}

We have modified the Tritt's model  to explain our observations ( Fig:1) in the following way. We propose that  in addition to the reorientation of the NH$_3^{+}$ group, the flip motion of the glycine group  (NH$_2$CH$_2$COOH) about the long axis of the GPI molecule may also contribute to the relaxation of  the protons particularly at higher temperatures. The relaxation will be dominated by the  NH$_3^{+}$ group reorientation at lower temperatures and at higher temperatures where $\omega \tau_{NH_3}\ll 1$, the 120$^0$ flip of the NH$_2$CH$_2$COOH group  could contribute significantly to the relaxation.  The 120$^0$ flip in GPI is supported by the second moment experiments carried out by Tritt {\it et.al.}~\cite{Tr98}.Similar observations  where the flip of the entire molecule contributing to the relaxation have been reported earlier in the literature~\cite{Ga84,Or71}. For example in the case of thiourea CS(NH$_2$)$_2$ the high temperature T$_1$ minimum  has been attributed to the  flip motion of NH$_2$ ~\cite{Or71}. We give below the expression for the proton NMR spin-lattice relaxation rate due to the combined effect of NH$_3^{+}$  reorientation and 120$^0$ flip  of NH$_3$CH$_2$COOH group along with spin-diffusion.
\begin{eqnarray}
\frac{1}{T_1} &=&  \frac{3}{8}D \left ( \frac{\tau_c}{1+\omega_0^{2}\tau_ c^{2}} +\frac{4\tau_c}{1+4\omega_0^{2}\tau_c^{2}}\right ) \nonumber \\
&&+\frac{2}{8}F \left ( \frac{\tau_i}{1+\omega_0^{2}\tau_i^{2}} +\frac{4\tau_i}{1+4\omega_0^{2}\tau_i^{2}}\right )
\label{Comb}
\end{eqnarray}
where $\tau_c$ and $\tau_i$ are the correlation times for the NH$_3^{+}$ group reorientation and NH$_3$CH$_2$COOH group flip motion respectively. Both the correlation times are assumed to follow the Arrhenius behaviour.  D and F represent the relaxation constants for the NH$_3^{+}$ group reorientation and NH$_3$CH$_2$COOH group flip motion respectively.  The factors $\frac{3}{8}$ and  $\frac{2}{8}$ come from spin-diffusion. The constant D is given as in Eqn:\ref{T1} and F is given by ~\cite{Or71},
\begin{equation}
F = \frac{27}{40}\left (\frac{\mu_0}{4\pi}\right )^2\frac{\gamma^4 \hbar^2}{R^6}\sin^2 \alpha
\end{equation}
where  $\alpha$ is the flip angle , R is the  inter -proton distance between the protons attached to the CH$_2$ group and the NH$_3$ group.  

\begin{figure}
\psfig{file=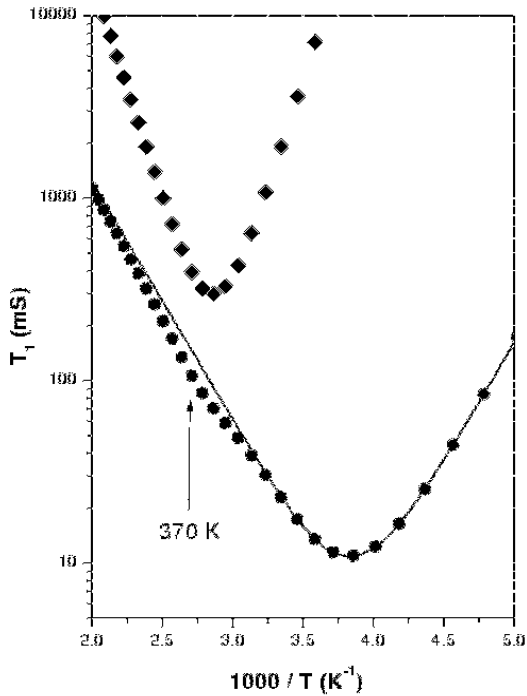,width=8cm}
\caption{Simulated T$_1$ vs 1000/T plot at 11.40 MHz using Equation:\ref{Comb}. Diamonds and line represent simulated T$_1$ data for 120$^0$ flip motion of glycine molecule and NH$_3$ group reorientation, respectively. Circles represent the data due to the combined motion of flip and reorientation.}
\end{figure}

We have done a least square fitting of the T$_1$ vs 1000/T data to Eqn: \ref{Comb} and the best fit is shown in Fig:1 and the best fit parameters obtained  at 11.40 MHz  and  23.56 MHz  are given in  the table \ref{RK model}. The correlation times obtained for the reorientation and flip motion compare well with those reported in the literature ~\cite{Ga84}. The activation energy of 26.8 kJ/mol for the  NH$_3^{+}$ group reorientation (which agrees with that of Tritt {\it et.al.}) is relatively high compared to that in the other glycine compounds like TGS, TGSe etc ~\cite{Bl66,Pi72,Sl82}. Tritt {\it et.al.}~\cite{Tr98} have attributed this high value to the relatively strong  hydrogen bonds in GPI compared to the other glycine systems. The high value of 45.6 kJ/mol for the activation energy of the flip of  NH$_3$CH$_2$COOH group is also to be expected and therefore this motion comes into play only at higher temperatures.  Similar values of activation energy for the flip motion have been reported in other systems like thiourea, hydraziniumnitrate etc ~\cite{Or71,Ga84}. The value of inter-proton distance in the NH$_3^{+}$ group calculated using the relaxation constant D is 1.69 $\AA$  and it matches well with the value of 1.70 $\AA$ obtained from neutron diffraction experiments in amino acids ~\cite{Ko76}. We have not observed any distinct T$_1$ minimum due to the flip motion. But a careful look at the spin-lattice relaxation behaviour (Fig:1) in the higher temperature region at 11.40 MHz , shows a weak indication around 370K.  As the flip motion is a high activation energy motion, the motional frequencies may not be very high and it is expected to show up significantly only at lower Larmour frequencies. Fig:2 shows the simulated T$_1$ vs 1000/T curve using the motional parameters from the table at 11.40 MHz for reorientational motion, flip motion and  the combined motion of reorientation and flip. The arrow  indicates the approximate temperature at which  T$_1$  minimum is expected  due to the flip motion. As the Larmour frequency increases the   T$_1$  minimum from the flip motion shifts to higher temperatures.  The data of Tritt {\it et.al.} was obtained at a higher Larmour frequency of 90 MHz  where the flip motion of the glycine group as a whole is not very effective and hence could be explained on the basis of NH$_3^{+}$ reorientation alone in the range of temperature studied.  $\omega_0 \tau_i$ at 11.40 MHz and 23.56 MHz  do not satisfy the condition $\omega_0 \tau_i \ll 1$ at higher temperatures and this explains the T$_1$ dispersion in the data at higher temperatures.

 It would be interesting to compare the behaviour of  T$_1$ near T$_c$ in GPI (T$_c$ = 224 K) and other glycine compounds which exhibit ferroelectric transition. In triglycine sulphate (TGS, T$_c$ = 320 K),selenate (TGSe, T$_c$ = 295 K), tetrafluoroberrylate (TGFBe, T$_c$ = 343 K), partially deuterated TGS (T$_c$ = 333 K) and potassium dihydrogen phosphate (KDP, T$_c$ = 123 K) the transition from the paraelectric phase to the ferroelectric phase has been found to be accompanied by a rather sudden decrease in the proton spin-lattice relaxation time ~\cite{Bl66,Bl68}. KDP is compared here due to its similarity with GPI in the direction of spontaneous polarisation.  In the TG systems  a sudden change in T$_1$ near the transition temeperature indicates a direct involvement of  protons in the  ferroelectric transition. In GPI also a  large isotopic effect (98 K) on T$_c$, points towards the essential role of the interphosphite hydrogen bonds in the phase transition mechanism. However  the T$_1$ varies smoothly with temperature near the ferroelectric transition temperature of 224 K ~\cite{Tr98} indicating that  the dynamics of the protons in the hygrogen bonds are fast for the NMR T$_1$ scale. This fast proton jump between the interphosphite hydrogen bond also could be the reason for a lower ferroelectric transition temperature in GPI compared to the other ferroelectrics mentioned above ~\cite{St97}. Vibrational spectroscopic studies show that normal H$_2$PO$_3^{1-}$ anions appear in the paraelectric phase and the ferroelectric phase, confirming that the protons in the interphosphite hydrogen bonds are dynamically disorderd in the time scale of vibrational spectroscopy ~\cite{Ba96}. These studies also suggest that the dynamical disorder of interphosphite hydrogen bonds is coupled to the motions of glycinium cations . The above statement is well supported by the splitting of some glycinium bands in the ferroelectric phase where the interphosphite hydrogen bonds settle to one of the positions in the time scale of vibrational spectroscopy. As stated above the motion of these protons appears to be too fast ($\omega_0 \tau_c \ll 1$) to be observed by the  NMR T$_1$ measurements.

\section{Conclusion}
We have observed a Larmour frequency dependence  in the spin-lattice relaxation time behaviour in GPI on the high temperature side of the T$_1$ minimum . Presence of inequivalent glycinium cations could not explain our observed results and an attempt has been made to explain this by invoking  the 120$^0$ flip motion of NH$_2$CH$_2$COOH group about it's long axis, in addition to the NH$_3^{+}$ reorientation . The motional parameters obtained match well with those obtained in the earlier studies. No significant change in T$_1$ behaviour could be observed at T$_c$.

We are grateful to Prof. H. L. Bhat and Ms.Deepthy. A. Pillai  for providing us the samples of Glycine phosphite used in these investigations.

\begin{table}
\caption{Motional parameters of GPI obtained using Equation:\ref{Comb}}
\begin{center}
\begin{tabular}{|c|l|} \hline
$\tau_{c0}$ &  5.2$\times10^{-14}\pm 3.6\times10^{-14}$ sec  \\ \hline
~
$\tau_{i0}$  & 1.9$\times10^{-15}\pm 0.72\times10^{-15}$sec   \\ \hline
~
E$_c$  & 26.8 $\pm$ 1.8 kJ/mol  \\ \hline
~
E$_i$ & 46 $\pm$ 0.16 kJ/mol \\ \hline
~
D & 1.25$\times 10^{10}$s$^{-2}$   \\ \hline
~
F & 0.04$\times 10^{10}$s$^{-2}$    \\ \hline
~
$\alpha$ & 120$\pm$7$^0$ \\ \hline
\end{tabular}
\end{center}
\label{RK model}
\end{table}

\end{document}